\documentclass[showpacs,twocolumn,prb]{revtex4}
\usepackage{graphicx}
\newcommand{\be}{\begin{equation}}
\newcommand{\ee}{\end{equation}}
\newcommand{\bea}{\begin{eqnarray}}
\newcommand{\eea}{\end{eqnarray}}
\newcommand{\bwt}{\begin{widetext}}
\newcommand{\ewt}{\end{widetext}}

\begin{document}
\title{Sum Rule Anomaly from Suppression of Inelastic Scattering in the Superconducting State}
\author{F. Marsiglio}
\affiliation{ Department of Physics, University of Alberta,
Edmonton, Alberta, Canada, T6G~2J1 \\
Department of Condensed Matter Physics, University of Geneva,
Geneva, Switzerland}
\begin{abstract}
In the conventional BCS description of a superconductor the kinetic
energy increases in the superconducting state. We describe the
observed decrease in kinetic energy by adopting a simple model of
electrons whose elastic scattering rate undergoes a sharp decrease
as the temperature is lowered below $T_c$. This phenomenology has
been suggested by other experiments, particularly microwave
conductivity. We find that such a decrease accounts for the observed
increase; a study of these different phenomena over a wide range of
high $T_c$ materials would confirm this correlation.
\end{abstract}

\date{\today}
\maketitle

\section{introduction}

Much of the focus in the high temperature cuprates has been on the
normal state. In fact, although we continue to probe the
superconducting state with finer probes and better samples, the
general conclusion seems to be that the state is BCS-like with
well-defined quasiparticles. Indications of this statement are from
photoemission \cite{kaminski00} and thermal conductivity
\cite{chiao00} studies on Bi2212 and YBCO, respectively. These and
many other studies are converging on the general notion that, while
the superconducting state has many conventional properties (albeit
with an order parameter with d-wave symmetry), the normal state is
in many ways anomalous.

However, optical studies over the last half dozen years have found
an anomalous behaviour in the optical sum rule \cite{kubo57}, that
appears in the superconducting state
\cite{basov99,molegraaf02,santander-syro03}. This sum rule relates
the weight of the entire oscillator strength to the bare plasma
frequency; a related sum rule pertains to a single band; then
optical processes involving transitions of excitations that belong
only to this band must be well-separated (even in principle) from
those that involve other bands. Alternatively, apparent violations
may occur because higher energy processes are ``left off the
accounting sheet'', so to speak.

The observations, whose interpretation remains under debate,
\cite{boris04,kuzmenko05,santander-syro05} can be summarized as
follows. First, recall the "single band" sum rule
\cite{maldague77,hirsch00,norman02}
\begin{equation}
\int_{0}^{+\infty }d\nu \mathop{\rm Re} \left[ \sigma_{\rm xx} (\nu
)\right] = {\pi e^{2} \over 4\hbar^2} \biggl\{ \frac{4}{N}\sum_{k}
{\partial^2\epsilon_k \over \partial k_x^2} n_{k}. \biggr\}%
\label{sumrule}
\end{equation}
where $\epsilon_k$ is the tight-binding dispersion and $n_k$ is the
single spin momentum distribution function. We define the quantity
in braces as $-<T_{\rm xx}>$. Note that in general the kinetic
energy is given by $<K> = \frac{2}{N}\sum_{k} \epsilon_k n_{k}$, and
only in the case of nearest neighbour hopping in two dimensions is
$<T_{\rm xx}> = <K>$. Nonetheless we will use this case to build
intuition concerning the behaviour of the sum rule. As explored in
Ref. \onlinecite{vandermarel03}, this characterization is reasonably
accurate even when beyond nearest neighbour hopping is included.

\begin{figure}[tp]
\begin{center}
\includegraphics[height=7.4in,width=4.0in]{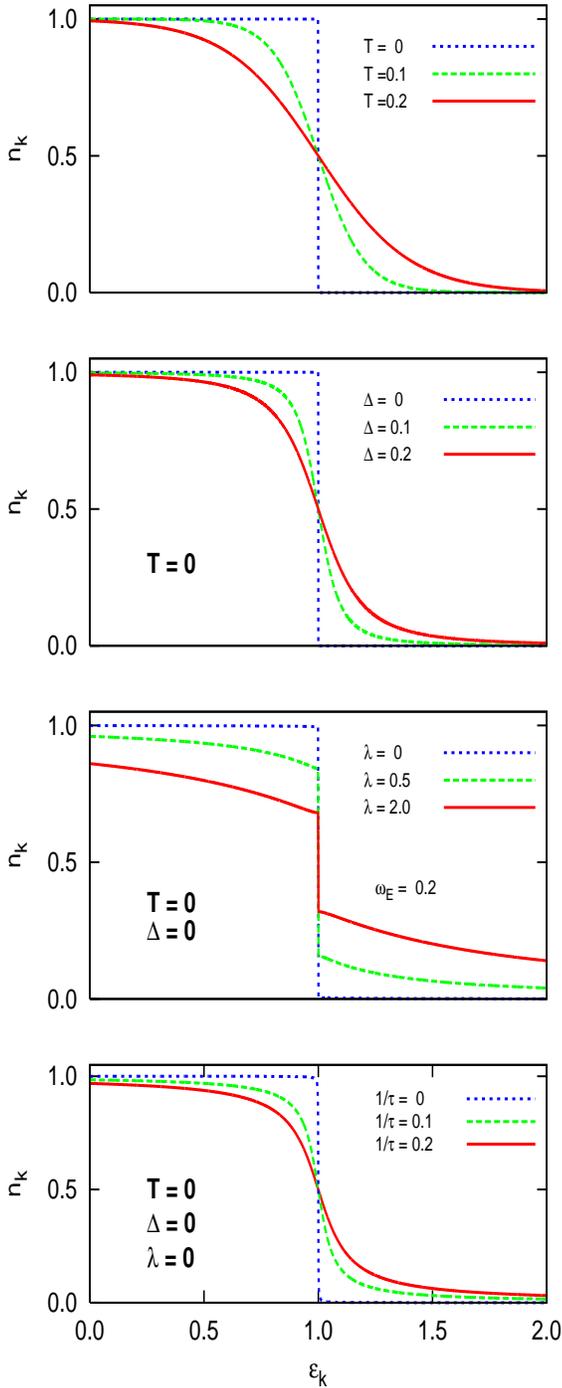} \caption{A summary of
various reasons for why the momentum distribution is broadened. In
(a) temperature is responsible for the broadening, while in (b) the
occurrence of superconductivity results in broadening, even at
$T=0$. In (c) and (d) inelastic scattering (as modelled by coupling
to an Einstein phonon  within the Migdal approximation) and elastic
scattering (as modelled by impurity scattering in the Born limit)
lead to the broadened distributions as shown.
 }
\end{center}
\end{figure}

The simplest situation in which one can understand Eq. \ref{sumrule}
is the case of non-interacting electrons; then, the momentum
distribution is simply given by the Fermi-Dirac function. At zero
temperature this is a step function at the Fermi energy, and all the
states with the lowest energy are occupied, while those with higher
kinetic energy are empty. This, of course minimizes the total
kinetic energy, and therefore maximizes the sum rule. As we raise
the temperature, states with higher kinetic energy become partially
occupied, while those with lower kinetic energy become partially
empty. This behaviour is illustrated in Fig. (1a). Therefore the RHS
of Eq. \ref{sumrule} decreases as temperature increases. The sum
rule (as determined by the RHS of Eq.\ref{sumrule}) is plotted in
Fig. 2 (uppermost red, solid curve) as a function of $T^2$. The
result is mostly linear in $T^2$, as one can discern from the
Sommerfeld expansion \cite{benfatto05}. However, note that this will
depend in general on the details of the band structure, filling,
etc.; careful inspection of Fig. (2) shows deviations from $T^2$
behaviour. This is partly caused here by the two dimensional van
Hove singularity. However, other band fillings, as we have verified,
also cause significant deviation from $T^2$ behaviour
\cite{benfatto05}.

What happens when the system goes superconducting? The momentum
distribution function is shown in Fig. (1b), for a gap of varying
size in the quasiparticle spectrum. Keep in mind that for an order
parameter with d-wave symmetry, the momentum distribution is no
longer a function of the band structure energy, $\epsilon_k$ alone.
For example, for a BCS order parameter with simple nearest neighbour
form, $\Delta_k = {\Delta \over 2} (\cos{k_x} - \cos{k_y})$, as $k$
varies from the $(0,0)$ to $(\pi,0)$, the magnitude of the order
parameter changes from zero to $\Delta$, On the other hand, as $k$
varies along the diagonal (from the bottom of the band to the top),
the order parameter is zero (and constant). In any event, Fig. (1b)
conveys the well-known fact that even at zero temperature, BCS-like
superconductivity {\em raises} the kinetic energy of the electrons.
The consequence of this for the electron kinetic energy is shown in
Fig. (2), for both an s-wave (green, dashed curve), and a d-wave
(blue, dotted curve) order parameter. As our discussion indicated,
the steady increase as the temperature is lowered is arrested at
$T_c$, and the {\em magnitude} of the kinetic energy in general {\em
decreases below $T_c$}. As expected, the effect is stronger for
s-wave than for d-wave symmetry.

\begin{figure}[tp]
\begin{center}
\includegraphics[height=2.4in,width=2.4in]{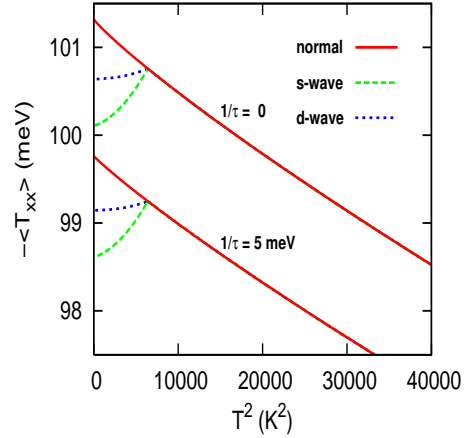}
\caption{The optical sum rule (RHS of Eq. (\ref{sumrule})) as a
function of temperature for the clean limit (upper curves) and with
elastic scattering (lower curves). Note that elastic scattering
mainly just shifts the curves. An s-wave order parameter (green,
dashed curve) leads to a more significant decrease in the sum rule
than a d-wave order parameter (blue, dotted curve). We used a
tight-binding model with nearest neighbour hopping only, as
described in the text. }
\end{center}
\end{figure}

Briefly, then, the experimental situation
\cite{molegraaf02,santander-syro03} is that, particularly for
underdoped and optimally doped samples, the conductivity sum rule
(i.e. the left-hand side (LHS) of Eq. \ref{sumrule}) indicates an
{\em increase} in the {\em magnitude} of the kinetic energy (RHS) in
the superconducting state. For the purposes of this work, we assume
this is the "experimental fact" and defer doubts and debates to
Refs. \onlinecite{boris04,kuzmenko05,santander-syro05} (a more
recent work is Ref. \onlinecite{carbone05}).

Interestingly, this very scenario was {\em predicted}
\cite{hirsch92} well before experiments with the required accuracy
were performed. The basis for this prediction was that the electron
(or hole) kinetic energy can, in principle, change upon entering the
superconducting state, so that the ``single band'' sum rule, Eq.
\ref{sumrule} would not remain constant. In particular spectral area
from relatively very high energy would change at the superconducting
transition. Since this spectral weight is not accounted for in the
single band sum rule of Eq. \ref{sumrule}, then there would be an
apparent violation. This was demonstrated to occur for the hole
mechanism \cite{hirsch89,hirsch89a} of superconductivity. In that
model the effective mass of the holes decreases in the
superconducting state; by pairing the holes are 'liberated' compared
to their mobility in the normal state. Partly this physics manifests
itself in an effective pairing interaction which has the peculiar
form obtained in Ref. \onlinecite{hirsch89a}. In this sense these
experimental results represent a "smoking gun" that point towards
the hole mechanism of superconductivity. Many details are worked out
in Ref. \onlinecite{hirsch00}.

However, other explanations have been put forth after the
experiments were performed. Many have focused on the impact of
superconductivity on the bosonic spectral function, i.e. on the
boson that is thought to mediate the pairing. For example, in Ref.
\onlinecite{norman02} a frequency-dependent scattering rate was
assumed, that underwent a significant reduction in the
superconducting state. A similar modelling was implemented in Ref.
\onlinecite{knigavko04} and Ref. \onlinecite{schachinger05}. In each
case, a change was put in by hand; in the first case they were
largely motivated by the Angular Resolved PhotoEmission Spectroscopy
(ARPES) results \cite{kaminski01}, while in the latter two the
appearance of a sharp neutron resonance mode \cite{rossat-mignod91}
provided the motivation for an abrupt change. In fact, in Ref.
\onlinecite{schachinger05} this latter scenario was explored in more
detail with models used for the spin fluctuations that have
described a number of other experiments.

The purpose of this paper is to see to what extent changes to a {\em
frequency-independent} scattering rate can reproduce the observed
sum-rule variation. The motivation is two-fold. First, it is always
helpful to simplify the description of the phenomenon as much as
possible, so that presumably only the essence remains. Secondly,
this follows the microwave analysis of Ref. \onlinecite{hosseini99},
and so would bring together observations from different experiments.
In that work (see also Ref. \onlinecite{bonn92}) a two-fluid
interpretation of their results led them to a frequency-independent
scattering rate whose value collapses in the superconducting state,
following a $T^4$ power law at low temperatures. This properly
accounted for the broad low temperature peak in the real part of the
microwave conductivity. Indeed, the original microwave experiments
\cite{nuss91} interpreted the presence of this peak in the same way;
both regarded the origin of this lifetime decrease as being due to a
"gapping" of some collective mode responsible for the normal state
scattering, but no longer functional in the superconducting state.
This is somewhat consistent with the scenario adopted by Norman and
P\'epin \cite{norman02} and Knigavko et al. \cite{knigavko04}. Here,
we wish to simplify the phenomenology that relates all of these
experimental findings: a simple number characterizes the single
electron scattering in the normal state (here this really is a
``characterization'', as the normal state is very likely not even a
Fermi Liquid). Thus linewidths are in general broad, as seen by
ARPES, for example, and collective modes of these electrons will
also be broad. A sharpening of the electron spectral function in the
superconducting state leads to quasiparticle-like peaks in the
ARPES, a collective mode that also sharpens, the prominent peak in
the microwave conductivity, and an apparent sum-rule anomaly below
$T_c$. We will focus on a description of the latter phenomenon. We
start with a discussion of the momentum distribution function in the
next section, where, in many cases an analytical form can be
obtained. We then explore the impact this has on the right-hand side
of the sum rule equation, Eq. (\ref{sumrule}), as a function of
temperature, order parameter symmetry, band structure, and elastic
scattering rate. Using the phenomenological temperature dependence
of the elastic scattering rate derived from the microwave
measurements we then illustrate the ``anomalous sum rule
behaviour'', as seen in experiment. We close with a summary and
implications for experiments.

\section{momentum distribution functions}

Returning to Fig. 1, we outline our calculations and explain the
other parts. In general, the momentum distribution function is given
by
\begin{equation}
n_k = {1 \over \beta} \sum_{m = -\infty}^{+\infty} G(k,i\omega_m)
e^{i\omega_m0^+} = {1 \over 2} + {1 \over \beta} \sum_{m =
-\infty}^{+\infty} {\rm Re} \ G(k,i\omega_m),
\label{momentum_distribution} %
\end{equation}
where $G(k,i\omega_m)$ is the (in principle) exact single electron
Green function with momentum $k$ and frequency $i\omega_m$. The
$i\omega_m$ are the Fermionic Matsubara frequencies,$i\omega_m = \pi
T (2m-1)$, where $m$ is an integer and $T$ is the temperature ($k_B
\equiv 1$). The inverse temperature, $\beta \equiv 1/T$. The Green
function $G$ is generally given by the Dyson equation,
\begin{equation}
G(k,z) = 1/\bigl[z - \epsilon_k - \Sigma(k,z)\bigr], %
\label{dyson}%
\end{equation}
where $z$ is anywhere in the complex plane, $\epsilon_k$ is the
non-interacting dispersion, and $\Sigma(k,z)$ is the electron self
energy. For the superconducting state, this equation holds when $G$
is understood to be the $G_{11}$ element of the $2\times 2$ Nambu
matrix \cite{schrieffer64}.

For the RHS of Eq. \ref{sumrule}, the calculation of the self energy
for a given model and approximation suffices, and these equations
are all that is required. For the LHS, the two-particle response is
required, and is much more difficult. Although we have computed the
LHS for various models in the past, this calculation is on less sure
footing, so we will show calculations of the RHS only. First, for
the non-interacting electron gas, $G_0(k,z) = 1/\bigl[z-\epsilon_k
\bigr]$, and the result from Eq. \ref{momentum_distribution} is the
Fermi function. This is what is plotted in Fig. 1a for the various
temperatures indicated, all in units where the bandwidth is 2, and
the chemical potential (hidden in $\epsilon_k$) is in the middle of
the band.

In Fig. (1b) we require the superconducting state. In this case the
relevant Nambu matrix element is
\begin{equation}
G_{11}(k,i\omega_m) = {i\omega_m Z(k,i\omega_m) + \epsilon_k \over
(i\omega_m)^2 Z^2(k,i\omega_m) - \phi^2(k,i\omega_m) -
\epsilon_k^2},
\label{nambu}%
\end{equation}
where the self energy is given by $\Sigma(k,i\omega_m) \equiv
i\omega_m \bigl[ 1 - Z(k,i\omega_m) \bigr]$, and an off-diagonal
self energy, $\phi^2(k,i\omega_m) \equiv
Z(k,i\omega_m)\Delta(k,i\omega_m)$, is also required. In BCS-like
theories it is the $\Delta(k,i\omega_m)$ that reduces to the
familiar gap parameter, $\Delta$. For the purposes of Fig. (1b) we
have assumed that something has given rise to superconductivity,
with a BCS order parameter, $\Delta_k$, whose values (constant in
$k$ for this figure) are indicated. Then $Z(k,i\omega_m) = 1$ and
$\Delta(k,i\omega_m) \rightarrow \Delta_k$. Using Eq. \ref{nambu},
Eq. \ref{momentum_distribution} yields the familiar
\cite{schrieffer64}
\begin{equation}
n_{k, s} = {1 \over 2} \biggl( 1 - {\epsilon_k \over E_k} \biggr),
\label{nks}%
\end{equation}
where $E_k \equiv \sqrt{\epsilon_k^2 + \Delta^2_k}$. A finite and
constant order parameter clearly gives rise to smearing at what was
the Fermi surface. Note that this is in the clean limit, and at zero
temperature. Nonetheless, the BCS ground state is usually considered
a Fermi Liquid, in the sense that well-defined quasiparticles (with
energy $E_k$) exist, even though there is no longer a discontinuity
at the chemical potential. It is clear that the onset of
superconductivity causes occupation of higher kinetic energy states,
just like raising the temperature does. If the onset of
superconductivity more than compensates the lowering of temperature,
than the trend which the kinetic energy was following with
temperature will reverse itself. Fig. 2 clearly indicates that this
is the case, though less so for an order parameter with d-wave
symmetry than one with s-wave symmetry.

In what other ways can higher kinetic energy states be occupied?
Refs. \onlinecite{norman02,knigavko04} used a coupling to a
phonon-like boson. Both applied the so-called Migdal approximation.
Using a single mode at Einstein frequency $\omega_E$, coupled to the
electrons with coupling constant $\lambda$, the self energy is given
by
\begin{equation}
\Sigma_{\rm in}(i\omega_m) = -i \lambda \omega_E \tan^{-1}\biggl(
{\omega_m \over \omega_E} \biggr).%
\label{migdal}%
\end{equation}
Then the momentum distribution function for inelastic scattering
becomes
\begin{equation}
n_{k, in} = {1 \over 2} - {1 \over \beta}\sum_m {\epsilon_k \over
\epsilon_k^2 + \bigl(\omega_m + \lambda \omega_E {\rm
tan}^{-1}\bigl( {\omega_m \over \omega_E} \bigr) \bigr)}.
\label{dist_in}%
\end{equation}
In the zero temperature limit, $\omega_m \rightarrow \omega$, a
continuous variable and the Matsubara sum becomes a continuous
integral. Then it is easy to see that a discontinuity remains at the
Fermi level, whose size is $1/(1 + \lambda)$. Calculations are shown
in Fig. (1c). Clearly increasing the coupling tends to push
electrons into higher kinetic energy states. If a lowering of the
electron-boson coupling were to accompany the onset of
superconductivity as the temperature is lowered, the savings in
kinetic energy could more than offset the squandering of kinetic
energy seen in Fig. 1b due to the superconductivity itself. This has
been amply discussed in Refs. \onlinecite{norman02,knigavko04}, and
will not be further addressed here.

The focus of this work is elastic scattering, for which we will
adopt the simple Born approximation. This will make everything we
compute look formally like impurity scattering (in the Born limit).
However, as Hosseini et al. \cite{hosseini99} pointed out, the
amount of impurities in their samples is small, and this issue has
been tested under controlled impurity addition, so we are really
modelling some inelastic process which, as discussed in the
introduction, seems to disappear when the material goes
superconducting. Elastic scattering is like static inelastic
scattering, i.e. in Eq. (\ref{migdal}) $\omega_E \rightarrow 0$
while $\lambda \omega_E \rightarrow 1/(\pi \tau)$, with $1/\tau
\equiv \Gamma$ some characteristic scattering rate. Then
$\Sigma_{\rm el}(k,i\omega_m) = -i {\Gamma \over 2}{\rm
sgn}(\omega_m)$, as expected for the Born approximation.
Substitution into Eq. \ref{momentum_distribution} then yields, for
elastic scattering,
\begin{equation}
n_{k, el} = {1 \over 2} - {1 \over \pi} {\rm Im} \ \psi \biggl( {1
\over 2} + {\bar{\Gamma} \over 2} + i\bar{\epsilon}_k \biggr),
\label{dist_el}%
\end{equation}
where $\psi(z)$ is the diagamma function, and $\bar{q} \equiv
q/(2\pi T)$. For $T=0$ one recovers the simple result
\begin{equation}
n_{k, el} = {1 \over \pi} \cot^{-1}\bigl({2 \epsilon \over \Gamma}
\bigr)   \phantom{aaaaa}{\rm (T = 0)}.
\label{dist_el_zerot}%
\end{equation}
This quantity is plotted in Fig. (1d). Note that at finite
temperature (shown below) more impurity scattering reinforces what a
higher temperature already does, i.e. it increases the kinetic
energy. Thus, it should be clear that a collapse of the scattering
rate will decrease the kinetic energy, and offset the trend evident
in Fig. (1b).

Finally, we wish to examine the distribution function in the
superconducting state. The quantities
\begin{eqnarray}
Z(k,i\omega_m) & = & 1 + {1 \over 2\tau} {1 \over \sqrt{\omega_m^2 +
\Delta_k^2}} \nonumber \\
\phi (k,i\omega_m) & = & \Delta_k + {1 \over 2 \tau} {\Delta_k \over
\sqrt{\omega_m^2 + \Delta_k^2}}%
\label{impurities}%
\end{eqnarray}
are entered into Eq. (\ref{nambu}), which is then inserted into Eq.
(\ref{momentum_distribution}). The result is
\begin{equation}
n_{k, els} = n_{0k, els} + \delta n_{k, els},%
\label{imp_express}%
\end{equation}
where
\begin{equation}
n_{0k, els} = {1 \over 2} - {\epsilon_k \over E_k}{2 \over \pi} {\rm
Im} \ \psi \biggl( {1 \over 2} + {\bar{\Gamma} \over 2} + i\bar{E}_k
\biggr),
\label{dist_els}%
\end{equation}
just like Eq. (\ref{dist_el}) except that $E_k$ replaces
$\epsilon_k$ at relevant places, and the additional piece,
\begin{widetext}
\begin{equation}
\delta n_{k, els} = -{2 \over \beta}\sum_{m=1}^\infty \biggl[
{\epsilon_k \over \epsilon_k^2 + \bigl(\sqrt{\omega_m^2 +
\Delta_k^2} + 1/(2\tau) \bigr)^2} - {\epsilon_k \over \epsilon_k^2 +
(\omega_m + 1/(2\tau))^2 + \Delta_k^2} \biggr]%
\label{tiny_piece}%
\end{equation}
\end{widetext}%
has to be evaluated numerically. Nonetheless, in the most extreme
case we looked at it represented less than $0.1\%$ of the total
result and can thus be safely ignored (we have included it in all
our displayed results). Eq. (\ref{dist_els}) is a nice compact form
that is also particularly simple at zero temperature:
\begin{equation}
n_{0k, els} ={1 \over 2}\biggl(1 - {\epsilon_k \over E_k}{2 \over
\pi} \tan^{-1}\bigl({2 E_k \over \Gamma}\bigr).
\label{dist_els_zero}%
\end{equation}
This form makes it clear that elastic scattering and the onset of
superconductivity both tend to broaden the distribution function, as
indicated by Fig. 1.

\section{the rhs}

Results for the RHS of Eq. (\ref{sumrule}) are most easily
determined by adopting a model density of states which is flat and
finite; one can then perform the integral straightforwardly
numerically (and in some cases analytically). However, this can't be
done when an order parameter with d-wave symmetry is present. Hence,
we have used a tight-binding band structure, and simply performed
the momentum sums (in two dimensions) in Eq. (\ref{sumrule}).

We begin with nearest-neighbour hopping only. Then results for the
clean limit were already displayed in Fig. 2. The additional curves
are shown for an elastic scattering rate $1/\tau = 5 $ meV. The
dashed, green (dotted, blue) curves are for s-wave (d-wave) order
parameters, respectively. Note that elastic scattering does almost
nothing to the results except for a constant shift downwards,
consistent with Fig. (1d). Even in the superconducting state, the
presence of elastic scattering simply gives rise to a constant shift
downwards.

To what extent does the "standard BCS" picture shown in Fig. 2
depend on the use of a tight binding model with nearest neighbour
hopping only? In particular, the hole- (and electron-) doped
cuprates have band structures that are often modelled with a
next-nearest neighbour contribution to the hopping. To explore this
issue in a little more depth, we define the relative difference
between the superconducting and normal state value of the sum rule
as:
\begin{equation}
\delta T_{i} \equiv {(-<T_{\rm xx}>_i) -\ \ (-<T_{\rm xx}>_N) \over
(-<T_{\rm xx}>_N)},%
\label{diff}%
\end{equation}
where $i = S,D$ refers to the superconducting state with s- or
d-wave symmetry, respectively, and the subscript $N$ refers to the
normal state, all at a given temperature. We use a similar
definition for the difference in the magnitude of the kinetic
energy, $\delta (-K_i)$. Inspection of Fig. 2 (where $\delta K_i =
\delta T_{i}$) indicates that this quantity is negative, with a
value of about $1\%$ for s-wave and about half that value for
d-wave. The important point is that it is negative. To see the sorts
of changes expected from a next-nearest neighbour amplitude and/or
change of electron density, we show in Fig. 3 a plot of $\delta
T_{i}$ vs. $t^\prime$ (curves), at zero temperature, for a band
structure, $\epsilon_k = -2t(\cos{k_xa} + \cos{k_ya}) + 4t^\prime
\cos{k_xa} \cos{k_ya}$. We use $t = 62.5$ meV, and allow $r \equiv
t^\prime/t$ to range from $-0.3$ to $0.3$, for two electron
densities, $n = 1$, and $n = 0.85$. We have taken $T_c = 80$ K, and
used BCS values for the order parameter as a function of
temperature; for s-wave, $2\Delta_0/k_BT_c = 3.5$, while for d-wave
$2\Delta_0/k_BT_c = 4.2$. For the temperature dependence we have
solved BCS equations in either case, with a small attractive on-site
potential for s-wave symmetry, and a small nearest neighbour
attractive potential for d-wave symmetry, on a square lattice.
Actually, we found that the temperature dependence is nearly the
same in both cases, and very well described by $\Delta(T)/\Delta_0 =
\sqrt{1 - (t^4 + t^3)/2}$, over the entire temperature range, where
$t \equiv T/T_c$.

Fig.~3 shows that there are minor variations with filling and band
structure. We have also shown the relative change in kinetic energy
(symbols) to illustrate that these are qualitatively the same as the
relative change in the sum rule, thus confirming van der Marel et
al.'s observation \cite{vandermarel03}. Note that for electron
doping one acquires the same results as for hole doping for negative
$t^\prime$. Another trend to note is that the relative value of the
effect due to superconductivity goes down as the overall scale of
the electronic band structure increases with respect to the
superconductivity energy scale ($80$ K $\approx 7$ meV).

\begin{figure}[tp]
\begin{center}
\includegraphics[height=6.4in,width=4.4in]{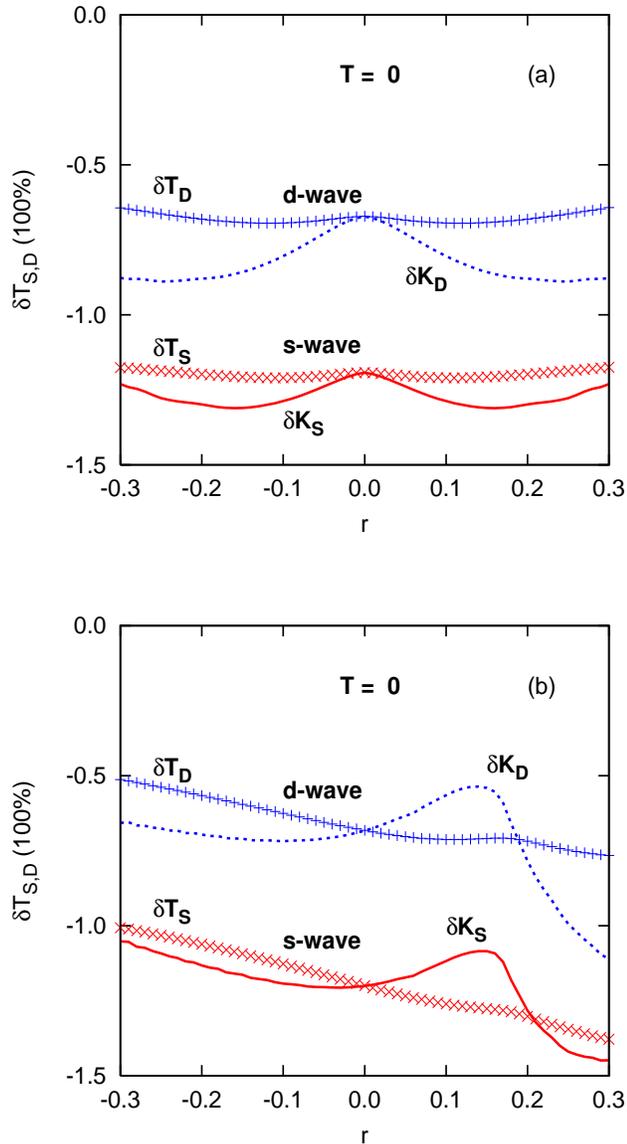}
\caption{ Plot of sum rule difference, Eq. (\ref{diff}) (curves) vs.
next-nearest neighbour hopping ($r \equiv t^\prime/t$) for (a)
half-filling, and (b) $n = 0.85$. The symbols are for the
corresponding difference in the magnitude of the kinetic energies.
They are in good qualitative agreement throughout the range of
interest.}
\end{center}
\end{figure}

\section{reversal of kinetic energy change at $T_c$}

Inspection of Fig.~2 (compare two red, solid curves) makes it clear
that a sudden drop in the scattering rate at $T_c$ will result in an
increase in spectral weight that can more than compensate the
decrease due to the opening of a superconductor gap. We simply adopt
the phenomenological observation of Hosseini et al.
\cite{hosseini99} and use the following model for the elastic
scattering rate, $\Gamma(T)$:
\begin{equation}
\Gamma(T) = \cases{ \Gamma_0 \phantom{aaaaaaaaaaaa} {\rm for}
\phantom{aaaa} T
> T_c, \cr
\Gamma_0 \bigl({T \over T_c}\bigr)^4 \phantom{aaaaaaa}{\rm for}
\phantom{aaaa} T < T_c. }%
\label{hosseini_model}%
\end{equation}
Note that we could include a linear temperature dependence in the
normal state, based on the linear resistivity; this would serve to
increase the (negative) slope in Fig.~2, in agreement with
observations. However, our attitude has been to adopt this simple
model to describe the superconducting state, given that we don't
really understand the normal state, so we will leave this as it is.
In the superconducting state, one could include a residual
scattering rate due to impurity scattering, as Hosseini et al.
\cite{hosseini99} have observed, even in their very clean crystals.
However, this is a minor effect, and also goes beyond the spirit of
the simple phenomenology that we are proposing here.

\begin{figure}[tp]
\begin{center}
\includegraphics[height=3.4in,width=3.0in]{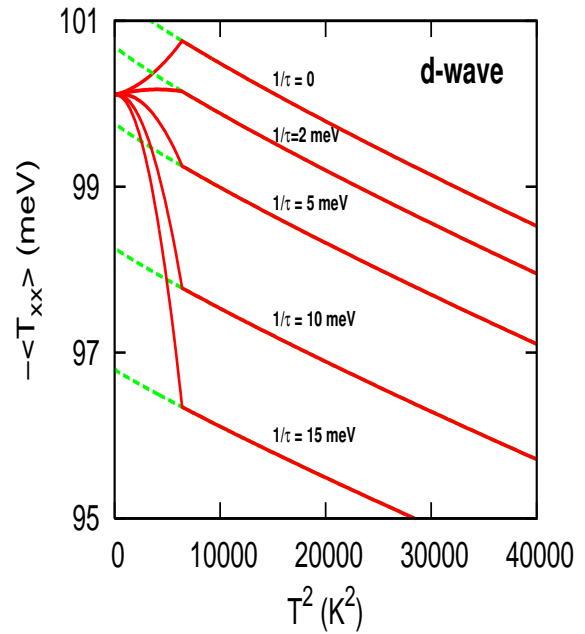}
\caption{Optical integral (RHS of Eq. (\ref{sumrule})) vs.
temperature, for various degrees of elastic scattering. Below $T_c$
we change the elastic scattering rate smoothly to zero (Eq.
(\ref{hosseini_model}), as the temperature is lowered. Further
details of the calculation are given in the text. }
\end{center}
\end{figure}

In Fig.~4 we show results for the sum rule with Eq.
(\ref{hosseini_model}) "put in by hand". Since we have assumed that
the crystals are completely free of impurities, the end point at
zero temperature is the same for all cases shown. It is difficult to
try to pin down the amount of inelastic scattering that gets
suppressed below $T_c$, based on this figure, because we can't use
the normal state result as a benchmark. Our slope is too low in
absolute value, and as already mentioned, the normal state is
undoubtedly {\em not} reasonably described by our simple approach.
Nonetheless, it is clear that the amount of inelastic scattering
that is suppressed in the superconducting state has to exceed a
minimum value so that the sum rule will exhibit anomalous behaviour.
As indicated, a small amount of inelastic scattering to begin with
(say, less than 2 meV for the parameters used in this figure), does
not make up for the decrease expected from conventional BCS theory.

\section{summary}

It is clear that Fig.~4 ``explains'' the anomalous optical sum rule.
To be precise it suggests that the origin of this anomaly is the
same mechanism that leads to the large coherence peak in the
microwave \cite{hosseini99} and to the other observations of sharply
increased coherence in the superconducting state \cite{johnson05}.
It is another matter, however, to claim that this increased
coherence is an intrinsic part of the mechanism. To our knowledge
the hole mechanism \cite{hirsch89,hirsch89a} is the only mechanism
for which this is the case in the plane. In other models, the
increased coherence generally arises due to a freezing of the
scattering mechanism, as originally suggested by Nuss et al.
\onlinecite{nuss91}.
Fig.~4 also indicates a source of non-universal
behaviour. If, for some materials or doping levels the loss in
inelastic scattering is small compared to the gap value, then more
conventional behaviour of the sum rule (eg. the uppermost curve in
Fig.~4) will prevail.

The virtue of this work is its simplicity. Without knowing the
details of the inelastic scattering process whose disappearance in
the superconducting state is responsible for a sharply decreased
scattering rate, we are able to understand the sum rule anomaly. Of
course this immediately implies correlations among the various
experiments. For example, should the anomaly not be present in
overdoped samples, then the microwave peak and neutron scattering
resonance should also be absent. Impurities should mask the effect,
since they presumably impact the superconducting state as well as
the normal state \cite{remark}, so it is important to have very
clean samples. We have adopted the $T^4$ dependence for the
scattering rate suggested in Ref. \onlinecite{hosseini99}, but the
qualitative anomaly will not depend on the detailed temperature
dependence, but rather on the overall magnitude of the change in
effective scattering rate.

\begin{acknowledgments}
I wish to thank Dirk van der Marel and Lara Benfatto for helpful
discussions. In addition the hospitality of the Department of
Condensed Matter Physics at the University of Geneva is greatly
appreciated. This work was supported in part by the Natural Sciences
and Engineering Research Council of Canada (NSERC), by ICORE
(Alberta), by the Canadian Institute for Advanced Research (CIAR),
and by the University of Geneva.
\end{acknowledgments}

\bibliographystyle{prb}

\begin{thebibliography}{1}
\bibitem{remark} Note that even in the case of an s-wave order
parameter, the impurity scattering is still present in the
superconducting state, and so the optical sum rule behaves as
illustrated in Fig.~2. Impurities become ``invisible'' in the
superconducting state only for some properties (like $T_c$, or the
density of states), but not for the kinetic energy.
\end{thebibliography}

\end{document}